\documentclass{article}
\usepackage{emulateapj}
\usepackage{psfig}
\usepackage{times}
\def\zycki{$\dot{\rm Z}$ycki}
\newcommand\fx{F_{\rm x}}
\newcommand\rozanska{R\'o$\dot{\rm z}$a\'nska }
\newcommand\lubinski{Lubi\'nski }
\def\simlt{\lower.5ex\hbox{$\; \buildrel < \over \sim \;$}}
\def\simgt{\lower.5ex\hbox{$\; \buildrel > \over \sim \;$}}
\def\sqr#1#2{{\vcenter{\hrule height.#2pt
      \hbox{\vrule width.#2pt height#1pt \kern#1pt
         \vrule width.#2pt}
      \hrule height.#2pt}}}

\righthead{X--ray irradiated disks}
\lefthead{Done \& Nayakshin}

\begin{document}
 
\title{Observational signatures of X--ray irradiated accretion disks}
\author{C. Done$^1$ \& S. Nayakshin$^{2,}$\altaffilmark{3}}
\altaffiltext{3}{National Research Council Associate}

\affil{$^1$ Department of Physics, University of Durham, South Road,
Durham, DH1 3LE, England; chris.done@durham.ac.uk}
\affil{$^2$ Laboratory for High Energy Astrophysics, NASA/Goddard
Space Flight Center,Greenbelt, MD20771, USA; serg@milkyway.gsfc.nasa.gov}

\begin{abstract}

Reflection of X--rays from cool material around a black hole is one of
the few observational diagnostics of the accretion flow
geometry. Models of this reflected spectrum generally assume that the
accretion disk can be characterized by material in a single ionization
state. However, several authors have recently stressed the importance
of the classic ionization instability for X--ray irradiated gas in
hydrostatic balance. This instability leads to a {\em discontinuous}
transition in the vertical structure of the disk, resulting in a hot
ionized skin above much cooler material. If the Compton temperature of
the skin is high then even iron is completely ionized, and the skin
does not produce any spectral features. These new models, where the
ionization structure of the disk is calculated self--consistently,
require an excessive amount of computing power and so are difficult to
use in directly fitting observed X--ray spectra. Instead, we invert
the problem by simulating X--ray spectra produced by the new
reflection models, and then fit these with the old, single zone
reflection models, to assess the extent to which the derived accretion
geometry depends on the reflection model used. We find that the single
zone ionization models can severely underestimate the covering
fraction of the ``cold'' material as seen from the X--ray source if the
optical depth in the ionized skin is of order unity, and that this can
produce an apparent correlation between the covering fraction and the
X--ray spectral index similar in nature to that reported by Zdziarski,
\lubinski and Smith (1999).

\end{abstract}


\section{INTRODUCTION}

Calculations of the reflected spectrum expected from X--ray
illumination of an accretion disk around a black hole are extremely
important. The reflected continuum and especially its associated iron
K$\alpha$ fluorescence line (see e.g.\ Basko, Sunyaev \& Titarchuk
1974; Lightman \& White 1988; George \& Fabian 1991; Matt, Perola \&
Piro 1991) are perhaps the only spectral features expected from
optically thick material in the vicinity of a black hole, so 
can be used to test theoretical models of
accretion in strong gravity (Fabian et al. 1989,
Dabrowski et al. 1997).

This is especially relevant since there is much current debate on the
structure of the accretion flow. The ``standard'' accretion disk models
of Shakura \& Sunyaev (1973: hereafter SS) are clearly
incomplete. They can produce UV and soft X--ray emission, but not the
hard X--ray power law out to 200 keV. This is observed to carry a
significant fraction of the luminosity in Active Galactic Nuclei (AGN)
and Galactic Black Hole Candidates (GBHC) in their low/hard
state. Something other than the SS disk is required. The two current
candidates are either magnetic reconnection regions above the
accretion disk, or that some part of the accretion flow is in an
alternative (non--disk) configuration.

The idea of magnetically powered active regions above a disk
(e.g. Galeev, Rosner \& Vaiana 1979) gained considerable plausibility
from the recent discovery that the accretion disk viscosity is driven
by a magnetic dynamo (Balbus \& Hawley 1991). Buoyancy could then make
the magnetic loops rise to the top of the disk, where the optical
depth is low so that the energy released in reconnection cannot
thermalize (e.g. di Matteo 1998). Such active regions could give the
hard X--ray spectra which are seen from AGN and Galactic Black Hole
Candidates (Haardt, Maraschi \& Ghisellini 1994; Stern et al. 1995;
Beloborodov 1999a,b).
Spectral transitions between the low/hard state and the high/soft
state in GBHC might then be associated with the disk becoming
radiation pressure dominated. This occurs at a few percent of the Eddington
accretion rate, roughly as observed in Cyg~X--1 and other
GBHCs. Perhaps magnetic buoyancy transfers more
energy out of the disk and into a corona in gas dominated disks
than in radiation pressure dominated ones (Nayakshin
1999a). Current (though still incomplete) 
MHD simulations do show strong magnetic fields
in the low density regions above and below the disk
(Hawley 2000), although they do not yet carry enough power to 
reproduce the observed low/hard state (Miller \& Stone 2000).

The idea of an alternative accretion flow became a serious
contender with the rediscovery of another stable solution of the
accretion flow equations for mass accretion rates below $\sim 10$ per
cent of Eddington. The standard SS disk model
derives the accretion flow structure in the limit where the
gravitational potential energy is radiated locally. This
assumption is relaxed in the Advection Dominated Accretion Flow (ADAF)
models, so that the energy can be transported radially (advected), as
well as radiated locally. There are two key (and currently
uncheckable) assumptions underlying this: firstly that the viscosity
mechanism gives the accretion energy to the protons (as opposed to
both protons and electrons) and secondly that the protons transfer
this energy to the electrons only via (rather slow) electron--ion
(Coulomb) collisions.  This gives an accretion flow which is optically
thin, geometrically thick and {\em hot}, with typical electron
temperatures of order 200 keV as observed (e.g. Narayan \& Yi
1995). This model provides a rather natural explanation for the
low/hard to high/soft spectral state transitions in GBHCs, as ADAFs
cannot exist at high mass accretion rates. The electron--ion collision
rate increases with the density of the flow, and so above a mass
accretion rate of $\sim 10$ per cent of Eddington (as long as the disk
viscosity parameter $\alpha\sim 0.2$; Esin, McClintock \& Narayan
1997, Quataert \& Narayan 1999) the 
electrons can drain most of the energy from the
protons. Advection is then unimportant as the electrons radiate the
gravitational energy released locally, and the flow collapses into an
SS disk.

In the one model, there is an optically thick disk extending down to
the last stable orbit at 3 Schwarzschild radii. In the alternative,
the inner disk is truncated, with the optically thick, cool flow
replaced by an optically thin, X--ray hot plasma. The reflected
spectrum gives an observational test between these two
scenarios. Firstly the amount of reflection should be different. A
disk with active regions above it should subtend a solid angle of
roughly $2\pi$, whereas a truncated disk with the X--ray hot plasma
filling the ``hole'' in the disk will subtend a rather smaller solid
angle. Secondly, for the SS disk there is material down to the last
stable orbit. Reflection features from the inner disk should then be
strongly smeared and skewed by the combination of special and general
relativistic effects expected in the vicinity of a black hole (Fabian
et al. 1989). This material is not present in the truncated disk
models, so they should have narrower lines.

The seminal ASCA observation of the AGN MCG--6--30--15 showed that the
line is so broad that accretion disk models require that the 
material extends down
to {\em at least} the last stable orbit in a Schwarzschild metric
(Tanaka et al. 1995), and that it 
subtends a solid angle of $\sim 2\pi$ with respect to the X--ray
source (Lee et al 1999).  
This very clear cut result gives strong support for the
existence of an SS accretion disk, although other interpretations 
in terms of optically thick, highly ionized 
cloudlets close to the black hole 
rather than a smooth disk flow may be possible (Karas et al 2000).

However, the line and reflected continuum seen from the GBHCs in their
low/hard state (where the X--ray continuum bears a remarkable
similarity to that from Seyfert 1 type AGN) are {\em not} the same as
that seen in MCG--6--30--15. The GBHC have reflection amplitudes which
are rather less than expected for a hard X--ray source above a disk
(e.g. Gierlinski et al. 1997; Done \&
\zycki\ 1999). The iron line does not generally show the extreme
smearing associated with material within 5 Schwarzschild radii,
although there is detectable broadening of the line showing that there
is cool, optically thick material within $\sim 10-20$ Schwarzschild
radii (\zycki, Done \& Smith 1997; 1998; 1999; Done \& \zycki\
1999). It is also emerging that while some Seyfert 1 AGN do have
extremely broadened iron lines like MCG--6--30--15 (e.g. NGC 3516;
Nandra et al. 1999), others look more like the GBHC in not having the
extreme skewed lines expected from the very inner disk (e.g. IC4329a;
Done, Madejski \& \zycki\ 2000), and that many do not show reflection
subtending the expected $2\pi$ solid angle (Smith \& Done 1996;
Zdziarski, \lubinski \& Smith 1999). 

Can this be interpreted as a strong evidence that the accretion flow
is not always given by an optically thick disk?  Such a conclusion is
plainly consistent with the data. Perhaps the objects with extreme
line profiles have high mass accretion rates, where the ADAF's cannot
exist. Perhaps these are the analogue of the GBHC in their high/soft
state (Done \& \zycki\ 1999; Zdziarski et al. 1999).  Alternatively,
Ross, Fabian \& Young (1999) argue that the
lack of strong reflection and extreme relativistic smearing does not
necessarily imply the absence of an inner disk, because the reflection
signature from the disk can be masked by complex ionization effects.
For completely ionized gas, there is no photo--electric absorption
opacity left, so the reflected spectrum is simply due to electron
scattering in the disk atmosphere. There is then no characteristic
iron line or edge feature from the inner disk, so the extreme
relativistically smeared components of the line are not present.

However, illumination of a disk with density given by the SS models
results in a power law distribution of ionization state with radius
(Ross \& Fabian 1993; Matt, Fabian \& Ross 1993). This does not fit
the data when the inner radii have a high enough ionization to mask
their reflected signature (\zycki\ et al., 1998; 1999; Done \& \zycki\
1999).  The key problem is that the observed reflected spectra from
the low/hard state GBHC are relatively neutral (\zycki\ et al., 1998;
1999; Done \& \zycki\ 1999).  Yet if the inner disk is extremely
ionized, then there should exist a transition region in which the disk
is cool enough for H-- and He--like iron to exist, before getting to the
low--to--moderate ionization material at larger radii. These species
give an unmistakable reflected signature, with very strong, highly
ionized line and edge features (Lightman \& White 1988; Ross \& Fabian
1993; Matt, Fabian \& Ross 1993ab; \zycki\ et al., 1994; Matt, Fabian
\& Ross 1996; Ross, Fabian \& Brandt 1996), which are simply not
present in the data. This leads to the conclusion that either the disk
really does subtend a small solid angle to the X--ray source, or that
its ionization state drops dramatically (as a step function rather
than a power law) from extreme ionization to relatively neutral (Done
\& \zycki\ 1999), or/and that the X--ray source does not illuminate
the inner disk (e.g. through the anisotropic illumination pattern
resulting from trans--relativistic outflow from magnetic reconnection:
Beloborodov 1999a,b).

This argument is based on reflection models which neglect the energy
change which can result from Compton scattering below $\sim 10-15$
keV, and assume that the
ionization state is constant with depth in the material.  In practice,
at high ionization states the electron temperature in the disk can be
high enough to cause appreciable Comptonization smearing 
of the resulting reflected
spectrum and line. Also, there is vertical ionization structure of the
illuminated material, the details of which depend on the
vertical density structure of the disk (Ross \& Fabian 1993; \zycki\
et al 1994; Matt, Fabian \& Ross 1996; Ross, Fabian \& Young 1999).
Ross, Fabian \& Young (1999) argue that these effects alone are
sufficient to hide the otherwise unmistakable signature of highly
ionized reflection from H-- and He-- like iron, but this has yet to be
tested against real data.

However, all these works use an {\em assumed} vertical density
structure. This is a very crude simplification, which can severely
misrepresent the reflection signature from the disk (Nayakshin 1999b;
Ross, Fabian \& Young 1999).  X--ray illumination {\em changes} the
density structure of the material. Material at the top of the disk is
heated by the illumination, and so can expand, lowering its
density. Deeper into the disk there is less X--ray heating, so the
material is cooler, and hence denser. The self--consistent density
structure is especially important as there is a thermal ionization
instability which affects X--ray illuminated material in pressure
balance (Field 1965; Krolik, McKee \& Tarter 1981; Kallman \& White
1989; Ko \& Kallman 1994; \rozanska \& Czerny 1996; \rozanska 1999;
Nayakshin, Kazanas \& Kallman 2000, hereafter NKK).  This can result 
in the ionization
state of the material changing very rapidly, with a highly ionized
skin forming on top of a mainly neutral disk. If the ionization state
of the skin is high enough to completely strip iron then it is almost
completely reflective, and forms no spectral features. The observed
reflected signature is then dominated by photons reflected from deeper
in the disk, where the material is mainly neutral. Although the
material is still expected to be less ionized at larger radii, the
meaning of less ionized is very different from that in the constant
density models -- the skin remains completely ionized with a
negligible amount of H-- and He--like iron, but its Thomson depth
decreases with radius (see Nayakshin 2000). The instability gives a
physical mechanism for a discontinuous transition in ionization state,
as required by the data.

The small observed solid angle subtended by the disk could then be an
artifact of fitting single zone ionization models to data from systems
with more complex ionization structure.  Ideally we would want to
reanalyze the data, fitting these X--ray illuminated disk reflection
models to derive the solid angle subtended by the reflecting
material. If these models give a worse fit to the data than the single
ionization models, or if they give a derived solid angle of the
reflector which is much less than $2\pi$ then the ``hole in the disk''
geometry inferred from the single zone ionization models is still
favored. If instead the complex ionization reflector models are a
better fit to the data and give a solid angle which is compatible with
$2\pi$ then the SS disk with magnetic coronal X--ray source is
probably correct, and the previous results which ruled out this
geometry are a model dependent interpretation of the data.

However, direct data fitting is prohibitively CPU intensive. The
models require simultaneous computation of the hydrostatic, ionization
and energy balance for the illuminated gas plus optically thick
radiation transfer and relativistic effects to determine the reflected
spectrum. A single model calculation requires $\approx 1$ day
of CPU on a 500 MHz workstation. Given this, we
have chosen to tackle the inverse problem.  Instead of fitting real
data with the complex ionization reflector code, we use the complex
ionization reflector code to synthesize a spectrum with the same
signal--to--noise and spectral resolution as given from typical {\em
GINGA} observations of the low/hard state spectra from Galactic Black
Hole Candidates. We then fit this model spectrum with the standard (and
computationally very fast) single zone ionization reflection models
which are widely used in X--ray spectral analysis. We can then
quantify the effect of using the single zone ionization reflection
models to describe a more complex ionization disk structure.

We show that the resulting spectra from a complex ionization
structure disk subtending a solid angle of $2\pi$ are well fit over
the 2--20 keV range by the single zone reflection models, but
that the presence of the ionized skin leads to an underestimate of the
covering fraction of the reflector. The SS disk models then cannot be
ruled out by the fact that single zone ionization reflector models
derive a solid angle of $\ll 2\pi$. The complex ionization structure
disks can also produce a correlation between the amount of reflection
as measured by the single zone ionization models and the spectral
index (as observed e.g. by Zdziarski et al., 1999) {\em without}
changing the solid angle subtended by the disk. We end by discussing
how we can observationally distinguish between the truncated disk
models and complex ionization structure.

\section{THE SPECTRAL MODEL}\label{sect:model}

We use the code of NKK to calculate the reflected spectrum from an
X--ray illuminated disk atmosphere.
This code computes the 
vertical structure assuming hydrostatic equilibrium, 
solving simultaneously for the density, ionization structure 
and energy balance of the illuminated gas, as well as doing the radiative
transfer for the continuum and emission lines. 
Locally, the amount of 
photo--ionization can be described by the parameter, $\Xi=\fx/(cP_{gas})$,
which is the ratio of the illuminating radiation pressure ($\fx/c$, where 
$\fx$ is the illuminating hard X--ray flux) 
to the gas pressure ($P_{gas}= 2.3 n_H kT$ 
acting outwards, where $n_H$ is the hydrogen
density and T is the temperature). This relates to the more usual
density ionization parameter $\xi=4\pi \fx/n_H$ by $\xi=2.3 \Xi
kT/4\pi c$. 

As results of NKK show, the vertical density structure of the
illuminated layer is rather complex (see also Nayakshin \& Kallman
2000), as the gas in radiative equilibrium with the illuminating
X--rays is subject to a thermal pressure ionization instability
(Krolik, McKee \& Tarter 1981).  In the context of the given problem,
the instability can be described as following. The X--ray heating
clearly depends on height, with the material at the top of the layer
being heated the most.  Because the material is in pressure balance,
the gas pressure is smallest at the top of the disk, so $\Xi$ is at
its maximum here. This corresponds to the uppermost stable branch of
the ionization equilibrium S--curve (see Fig. 1 in NKK), with low
density material which is highly or completely ionized. Atomic cooling
is negligible so the temperature is a fraction of the local Compton
temperature (see Krolik et al., 1981 and Nayakshin 2000a); Compton
heating is balanced by bremsstrahlung and Compton cooling.  Going
deeper down into the disk, the X--ray heating decreases with optical
depth to electron scattering, $\tau_h$, and the gas pressure
increases. Hence, both the temperature and ionization parameter $\Xi$
decrease. However, there is then a turning point on the S--curve
(point c in Fig. 1 of NKK). At this point the temperature and density
is such that the rapidly increasing bremsstrahlung cooling causes the
temperature to dramatically decrease. This pulls the gas pressure
down, but hydrostatic equilibrium requires that the gas pressure must
monotonically increase as a function of depth into the disk.  The only
way for the pressure to increase in a region with rapidly decreasing
temperature is to rapidly increase the density, but this pushes up the
cooling still further, and so decreases the temperature.  Ionic and
atomic species can now exist, so line transitions give a yet further
increase in the cooling. Eventually this stabilizes onto the bottom
part of the S curve, where the X--ray heating is balanced primarily by
line cooling from low temperature, high density material. This results
in an almost discontinuous transition from a highly ionized, hot and
relatively tenuous skin to mainly neutral, cool and relatively dense
material over a very small ($\Delta\tau_h \sim 10^{-3}$ -- see
Appendix in NKK) change in optical depth at $\Xi=\Xi^*\sim$ few.

The optical depth where this instability occurs plus the Compton
temperature (which depends on the illuminating power law as well as
the local thermal disk spectrum) are the main determinants of the
resulting reflected spectrum.  The majority of the reflected flux
arises from the hot layer if its optical depth $\tau_h> 1$. The
reflected spectrum then contains the imprints of the ionized material
which can be either highly ionized (where iron is mainly He-- or
H--like), or extremely ionized (where even iron is completely stripped
of bound electrons) depending on the equilibrium Compton
temperature (Nayakshin \& Kallman 2000). 
For smaller optical depths, $\tau_h < 1$, most of the
reflection occurs in material below the ionization instability point,
so the reflected spectrum contains the signatures of the cool material
at low ionization states.  NKK define a dimensionless parameter $A$,
which is essentially the ratio between the gravity force and that from
the X--ray radiation pressure ($1.2 n_H \sigma_t \fx/c$). Where gravity
is dominant ($A>>1$) then the vertical disk structure is close to that
of an un--illuminated disk (assumed here for simplicity to be the
Gaussian density profile expected from a gas pressure dominated disk:
Shakura \& Sunyaev 1973) and the optical depth of the hot, ionized
layer is very small.  Conversely, where $A<<1$ then the illumination
determines much of the vertical structure, and the optical depth of
the hot layer is large.  NKK give a simple analytic approximation to
this hot layer optical depth of $\tau_h\sim (\Xi^* A)^{-1}$ (see also
Kallman \& White 1989; Nayakshin 2000 for more accurate analytical
estimates of $\tau_h$).

Given that the observations of flat spectrum AGN and low/hard state
GBHC seem to show small but not negligible amounts of low to moderate
ionization reflection, then the parameters which are most likely to
reproduce this are $\tau_h\sim 1$, which corresponds to $A\sim 0.3$
for hard spectra. If the Compton temperature in the upper, ionized
skin is high enough then the reflected spectrum from this will be
featureless, and so not detected in X--ray spectral fitting. 
The observable reflection signature will then be given only by the
material below the ionization instability point, where the incoming
flux has been decreased by scattering in the hot layer. This 
reflected continuum then has to escape out through the hot layer, 
so scattering again reduces its amplitude and 
introduces some broadening to the reflected features.

\begin{figure*}[!h]
\centerline{\psfig{file=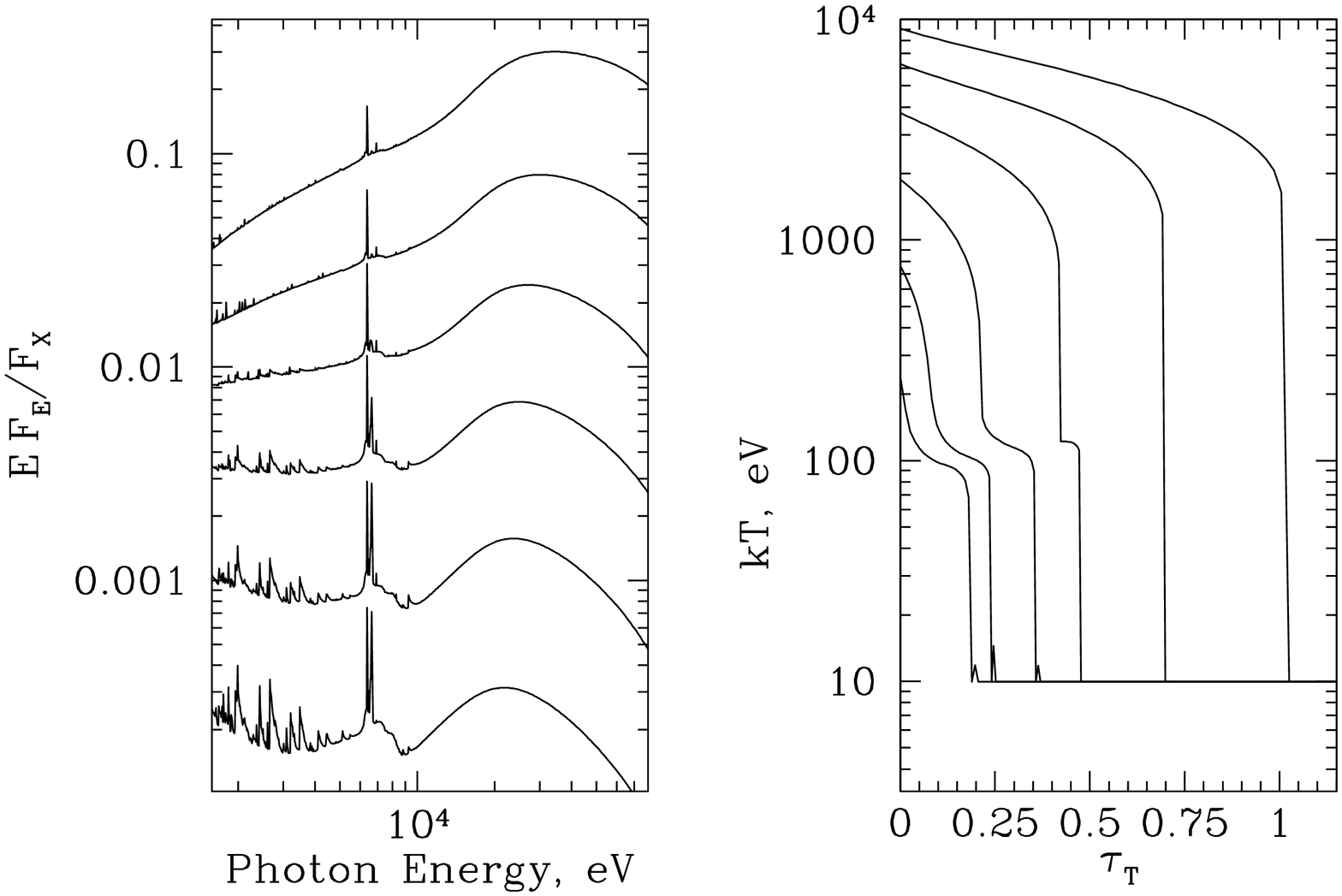,width=.9\textwidth,angle=-90}}
\caption{Reflected spectra (left) and temperature profiles (right)
for 6 different spectral indices ($\Gamma = $ 1.5, 1.7 ... 2.5)
calculated with the NKK code for a fixed gravity parameter $A= 0.3$.
Note that for harder spectra, the gas temperature is higher and the
ionized skin is thicker. This gives the apparent
correlation of $\Gamma$ and $\Omega/2\pi$ shown in Figure 2}
\label{fig:spectr_tempr}
\end{figure*}

A truly self--consistent model should uniquely determine the
illuminating flux at the surface of the disk as a function of the
system parameters (mass, radius and mass accretion rate through the
disk). Since the X--ray energy generation mechanism for magnetic
flares is not yet understood in detail then such a model is not yet
available.  Instead we {\it assume} a given X--ray illuminating flux
in order to demonstrate the physical behavior of a disk atmosphere
irradiated by magnetic flares. We calculate the angle averaged upward
reflected spectra (corresponding to an inclination of $\sim 60^\circ$)
resulting from a disk at $R= 10 R_S$ ($R_S=2GM/c^2$, the Schwarzschild
radius), with an accretion rate of $\dot{m}=10^{-3}\dot{m}_{Edd}$ onto
a super--massive black hole of $10^8 {\rm M}_\odot$. The large black
hole mass means that we avoid disk photons potentially
contributing to the observed bandpass. We assume a power law
illuminating continuum with exponential cutoff at 200 keV. The
illumination parameter is fixed at $A=0.3$, and the resulting angle
averaged upward reflected spectra (so corresponding to an inclination
of $\sim 60^\circ$) are calculated for photon indices of $\Gamma=1.5,
1.7, 1.9, 2.1, 2.3$ and $2.5$. We do not include any general or
special relativistic shifts for simplicity.  Figure
\ref{fig:spectr_tempr} shows the reflected spectra (without the
illuminating power law) and the gas temperature profiles for the 6
different values of $\Gamma$. 

\section{Overall 2--20 keV Spectral Shape}\label{sect:fits}

We simulate a 10ks observation with the GINGA X--ray satellite, using
the {\tt fakeit} command in XSPEC (Arnaud 1998), adding
0.5 per cent systematic errors to the results since it seems unlikely
that any satellite will be calibrated to better than this limit.  We
compare these simulated spectra with the angle dependent reflection
code of Magdziarz \& Zdziarski (1995), with ion populations calculated
as in Done et al. (1992), implemented in the {\tt pexriv} model in
XSPEC. This calculates the continuum reflection for an input
spectrum which is a power law with exponential cutoff. It does not
include the associated iron line, so this is put in as a narrow 
(intrinsic width fixed at $0.1$ keV) Gaussian line
of free normalization, and energy (except that this latter is
constrained to be between 5.5 and 7.5 keV, as expected for iron with
some shifts allowed for Doppler and gravitational effects). We assume
solar abundances (Morrison \& McCammon 1983) and an inclination of
$60^\circ$. We also include a free absorbing column in our fits so as
to match as closely as possible the uncertainties which arise in
fitting real data.

\begin{figure*}
\centerline{\psfig{file=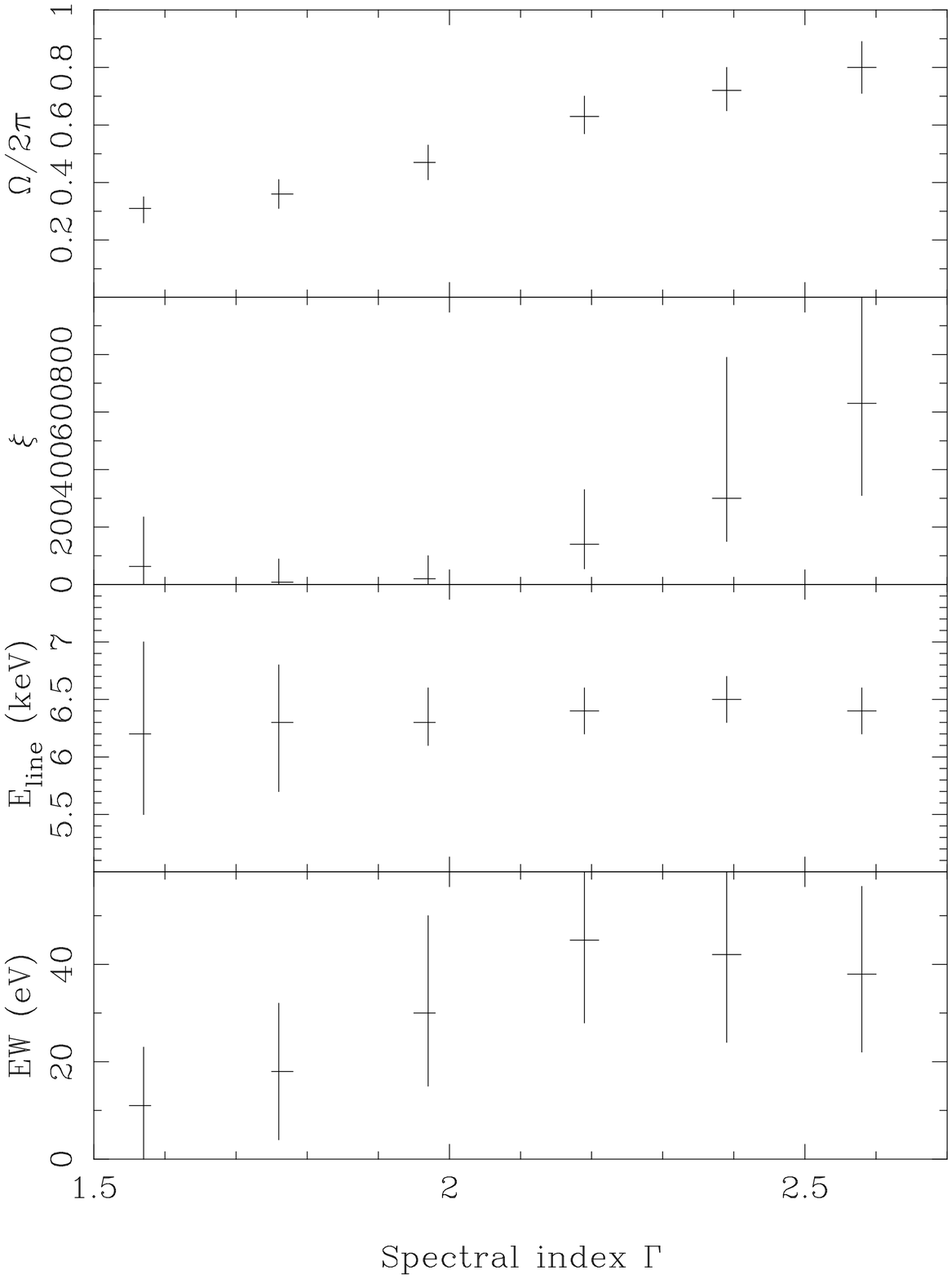,width=.7\textwidth}}
\caption{
The complex ionized disk models of Nayakshin et al. (2000) for $A=0.3$
as seen in a $10^4$ second GINGA exposure are fit to a model of a
power law (with exponential cutoff) and a single zone ionized
reflection model with separate iron line.  The best fit reflected
spectral parameters are clearly correlated with the parameters of the
illuminating spectrum. The amount of reflection $\Omega/2\pi$
increases with spectral steepness, as does its ionization state. The
iron line energy and equivalent width errors are too large to
independently constrain the ionization state}
\label{fig:2}
\end{figure*}

Figure \ref{fig:2} shows the spectral index and amount of reflection
inferred from fitting the simulated datasets with the {\tt pexriv}
model, with full results given in Table 1. There is a clear
correlation of the amount of reflection and its
ionization state with spectral index. 
For hard spectra the ionization of the detected
reflected spectrum is low,
and the amount of reflection is much less than unity.  This is because
the Compton temperature of the upper layer of the disk is high, its
ionization is extreme and it is Thomson thick ($\tau_h\simeq 1$, see
Fig. \ref{fig:spectr_tempr}). This produces a reflected spectrum which
is mostly from electron scattering in the skin, which has few spectral
features and cannot easily be identified as reflected flux. The
ionization drops dramatically at the depth where the ionization
instability sets in, so there is also a small reflected component from
low ionization material which is fit by the {\tt pexriv}
model. However, this itself has to escape out through the skin, and so
some fraction is Compton scattered, reducing the line and edge
features from the low ionization reflected spectrum, but keeping its
overall shape. 

For steeper spectra the Compton temperature of the upper
layers drops, so that its ionization is not so extreme, leading to
strong spectral features from the ionized upper layer, plus the skin
is actually thinner (see Fig. 1), and hence most of the reflected flux
is easily identified. Figure \ref{fig:2} also shows the detected line
equivalent width. It is generally low, but consistent with the
theoretically expected equivalent width for the given ionization state
(assuming that resonant trapping is not important) and spectral index,
scaled by the observed solid angle (\zycki\ \& Czerny 1994; George \&
Fabian 1991).

\begin{deluxetable}{lcccccccc}
 \footnotesize \tablewidth{0pt} \tablecaption{Simulated GINGA spectra
 from the illuminated disk models of Nayakshin et al., (2000) fit to
 simple reflection models (with assumed solar abundances and
 inclination of $60^\circ$). Errors are calculated for
 $\Delta\chi^2=2.7$} \tablehead{
\colhead{Intrinsic $\Gamma$} &
                     \colhead{Fit $\Gamma$} &
		     \colhead{$E_{\rm cut}$\tablenotemark{a}} &
                     \colhead{Norm\tablenotemark{b}} & 
                     \colhead{$\Omega/2\pi$\tablenotemark{c}} &
                     \colhead{$\xi$\tablenotemark{d} }  &
	             \colhead{Line Energy\tablenotemark{e}} &
                     \colhead{Line Eqw (eV)} &
\colhead{$\chi^2_\nu$} \nl
}

\startdata
$1.5$ & $1.57_{-0.02}^{+0.01}$ & $>1100$ & $2.3$ & $0.31_{-0.05}^{+0.04}$ &
$63_{-62.9}^{+172}$ & $6.2_{-0.7}^{+0.8}$ & $11_{-11}^{+12}$ &
$12.9/22$ \nl 
$1.7$ & $1.76_{-0.01}^{+0.02}$ & $>1000$ & $5.4$ & $0.36\pm 0.05$ &
$8.5_{-8.4}^{+80}$ & $6.3_{-0.6}^{+0.5}$ & $18\pm 14$ &
$16.5/22$ \nl 
$1.9$ & $1.97_{-0.02}^{+0.01}$ & $>1000$ & $7.2$ & $0.47\pm 0.06$ &
$20_{-19.3}^{+80}$ & $6.3_{-0.2}^{+0.3}$ & $30_{-15}^{+20}$ &
$18.1/22$ \nl 
$2.1$ & $2.19 \pm 0.02$ & $>1200$ & $10.0$ & $0.63_{-0.06}^{+0.07}$ &
$ 140_{-85}^{+190}$ & $6.4\pm 0.2$ & $45_{-17}^{+16}$ &
$25.2/22$\nl 
$2.3$ & $2.39\pm 0.02$ &  $>1200$ & $13.2$ & $0.72^{+0.08}_{-0.07}$ & 
$300_{-150}^{+490}$ & $6.5\pm 0.2$ & $42_{-18}^{+19}$ & $25.2/22$\nl  
$2.5$ & $2.58\pm 0.02$ & $>800$ & $17.1$ & $0.8\pm 0.09$ & 
$630_{-320}^{+800}$ & $6.4\pm 0.2$ & $38^{+18}_{-16}$ & $10.9/22$ 
\nl

\enddata
\tablenotetext{a} {Exponential cutoff energy (keV)}
\tablenotetext{b}{Normalization at 1 keV}
\tablenotetext{c}{Solid angle subtended by the reprocessor with respect to the
X--ray source.}
\tablenotetext{d}{Ionization parameter of the reprocessor}
\tablenotetext{e}{Line energy (keV) constrained between 5 and 7 keV for
stability of the fit}
\end{deluxetable}

The high energy cutoff is never detected (see Table 1), despite the
fact that an exponential roll--over temperature at 200 keV gives a 10
per cent effect at 20 keV. This is probably due to the 
weak line and edge features compared to the Compton hump 
in the complex models due to Compton
scattering of the low ionization reflected signature in the skin. 
Therefore, for a given
reflected {\em continuum}, the models of NKK always have less of the
absorption edge than the single ionization zone models (see also Ross,
Fabian \& Young 1999 who came to similar conclusions for exponential
density profiles).

Hence, when using the single zone models, by fixing the depth of the
photo--absorption edge (with the edge energy fixing the ionization
state), one in fact picks a particular value for $R$ already. This
value is too small to reproduce the continuum reflection rise of our
models shown in Fig. \ref{fig:spectr_tempr}, but the fitting
compensates for this by increasing the exponential cutoff energy.
However, we caution that an exponential cutoff is much less sharp than
the roll--over in a true Comptonized spectrum, so this aspect of the
modeling need not be reproduced in detail by real illuminated disks.

All the fits are statistically very good, with $\chi^2\le 25$ for 22
degrees of freedom. However, there should be some observable features
from these disk models with structured ionization which distinguish
them from simple single ionization state models. Figure 3 shows the
total spectrum and reprocessed components from the $\Gamma=1.5$ flat
spectrum simulation, together with the best fit model using the {\tt
pexriv} simple (single zone) ionization reflected spectrum.  Plainly
the absorbed power law and its reflected ({\tt pexriv}) component give
an excellent fit to the total spectrum derived from the complex
ionization state models, but the individual components are not at all
well modeled. The true reflected flux is dramatically underestimated
and the derived ``intrinsic'' spectrum is somewhat steeper. These two
together would give a rather steeper total spectrum, but this is
compensated for by the lack of high energy cutoff in the fitted
spectrum. With broad band data, extending significantly above 20 keV
then there should be significant residuals from fitting true complex
ionization spectra with the {\tt pexriv} code.

\begin{figure*}
\centerline{\psfig{file=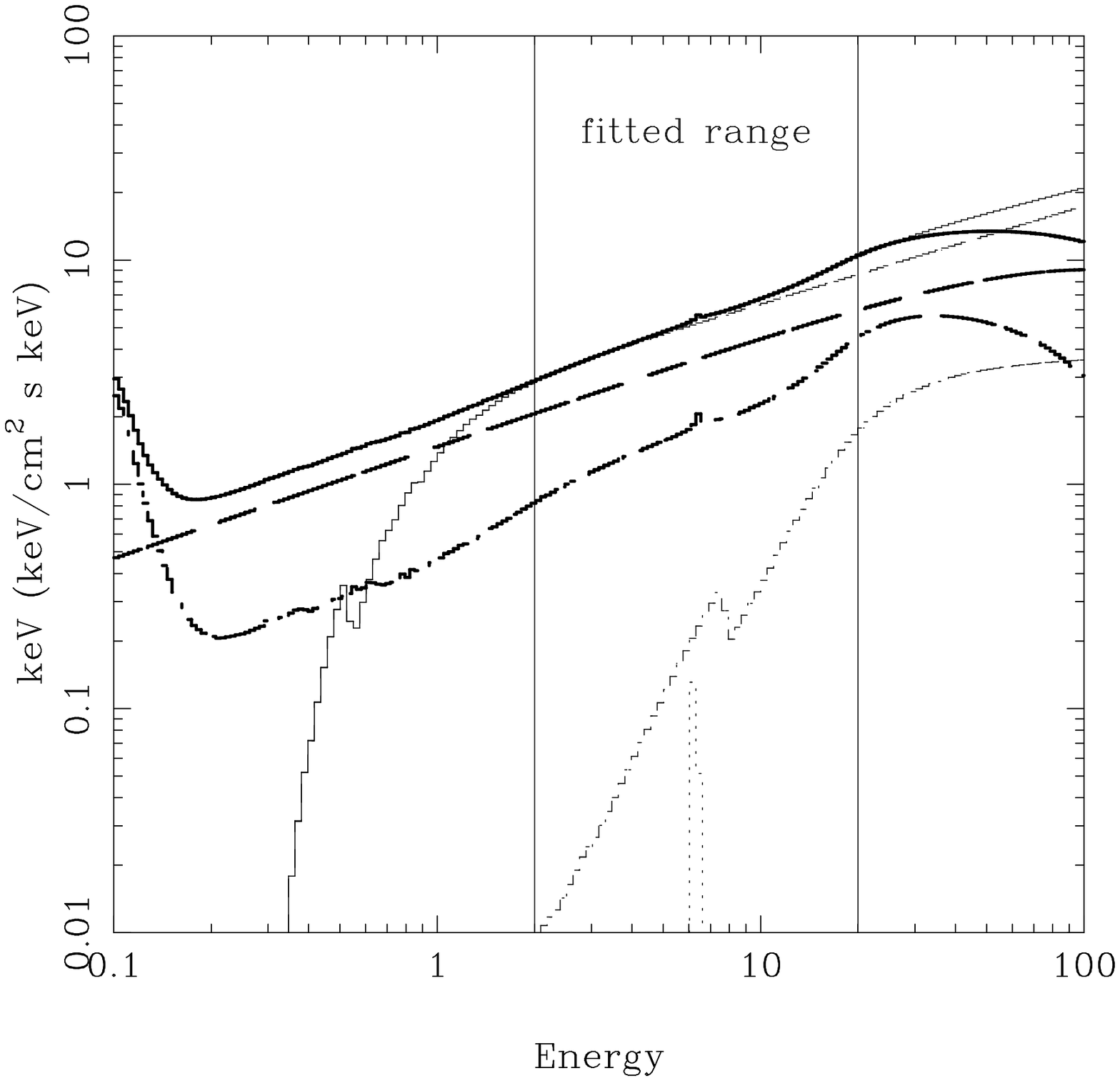,width=.7\textwidth}}
\caption{
The heavy lines show the complex ionization model spectrum for
$\Gamma=1.5$. The long dashed line is the intrinsic illuminating
spectrum, while the dash--dot line shows the reflected flux and the
solid line is their total. The thin lines show the best fit single
zone ionization model components, again with the long dashed line
giving the derived illuminating continuum, the dash--dot line giving
the reflected continuum. The doted line is the separate Gaussian line
and the solid line is the total spectrum. The bets fit single zone
ionization reflected continuum dramatically underestimates the true
reflected spectrum, but the total spectra over the fitted 2--20 keV
range are almost identical}
\label{fig:3}
\end{figure*}

Even in the 2--20 keV band it is plain that there is excess reflected
flux at low energies in the stratified ionization model which is not
matched by the single zone ionization state code. This excess is at
the 10 per cent level, and should easily be detectable. We include a
broad soft excess component in the fit to mimic this, using the {\tt
compst} (Sunyaev and Titarchuk 1980) model in XSPEC. This leads to a
significant reduction in $\chi^2$ for the flat spectra (see table
2). The resulting reduced $\chi^2$ values are typically very much less
than unity because the model is sometimes an adequate fit to the
simulated data {\em without} systematic errors (photon counting errors
are of the order of 0.05 per cent!). Error ranges are then derived
from $\Delta\chi^2/\chi^2_\nu=F=2.7$, rather than the standard
$\Delta\chi^2=2.7$ which assumes that the fit has $\chi^2_\nu\sim 1$.
Several of the parameters are tightly correlated, so error ranges on a
single parameter are somewhat misleading.

Figure 4 shows the resulting fit to the simulated $\Gamma=1.5$
spectrum. Comparison with Figure 3 shows that the broad soft excess
component is used to describe the highly ionized reflected flux at low
energies, while the fitted reflected flux is still dramatically underestimated.

\begin{deluxetable}{lcccccccc}
 \footnotesize \tablewidth{0pt} \tablecaption{Simulated GINGA spectra
 from the illuminated disk models of Nayakshin et al. (2000) fit to
 simple reflection models and a soft excess component. Errors are
 calculated for $\Delta\chi^2=2.7\times \chi^2_\nu$} \tablehead{
\colhead{Intrinsic $\Gamma$} &
                     \colhead{Fit $\Gamma$} &
                     \colhead{Norm} & 
	             \colhead{$kT$} &
                     \colhead{$\tau$} &
                     \colhead{Compst Norm} &
                     \colhead{$\Omega/2\pi$\tablenotemark{a}} &
                     \colhead{$\xi$\tablenotemark{b} }  &
\colhead{$\chi^2_\nu$} \nl
}

\startdata

$1.5$ & $1.37^{+0.09}_{-0.08}$ & $1.4$ & $1.9^{+0.6}_{-0.2}$ &
$15.5^{+1.5}_{-2.5}$ & $0.6$ & $0.12^{+0.13}_{-0.09}$ &
$60^{+2000}_{-56}$ & $1.00/19$ \nl
$1.7$ & $1.55\pm 0.1$ & $3.3$ & $1.9^{+1.3}_{-0.2}$ &
$14\pm 4$ & $1.6$ & $0.14^{+0.39}_{-0.14}$ & $4^{+\infty}_{-4}$ &
$3.85/19$ \nl
$1.9$ & $1.75_{-0.11}^{+0.08}$ & $4.2$ & $2.1_{-0.3}^{+1.4}$ & 
$11.6_{-2.7}^{+2.6}$ & $2.6$ & $0.23_{-0.11}^{+0.38}$ &
$50_{-46}^{+930}$ & $4.05/19$\nl
$2.1$ & $1.90^{+0.06}_{-0.04}$ & $5.4$ & $1.8^{+0.7}_{-0.3}$ &
$11^{+1}_{-3}$ & $4.0$ & $0.22^{+0.18}_{-0.08}$ & $900^{+3000}_{700}$ &
$3.66/19$ \nl
$2.3$ & $2.12^{+0.06}_{-0.08}$ & $7.5$ & $1.5^{+0.6}_{-0.2}$ &
$11.5^{+1.3}_{-1.5}$ & $4.9$ & $0.2^{+0.16}_{-0.06}$ &
$3000^{+16000}_{-1000}$ & $5.36/19$ \nl
$2.5$& $2.40\pm 0.04$ & $11.9$ & $1.3^{+0.5}_{-0.3}$ & 
$12\pm 3$ & $3.7$ & $0.37\pm 0.07$ & $3000^{+3000}_{-1000}$ & $2.50/19$ \nl

\enddata

\tablenotetext{a}{Solid angle subtended by the reprocessor with respect to the
X--ray source (assuming solar abundances and inclination of $60^\circ$)}
\tablenotetext{b}{Ionization parameter of the reprocessor}
\end{deluxetable}

\begin{figure*}
\centerline{\psfig{file=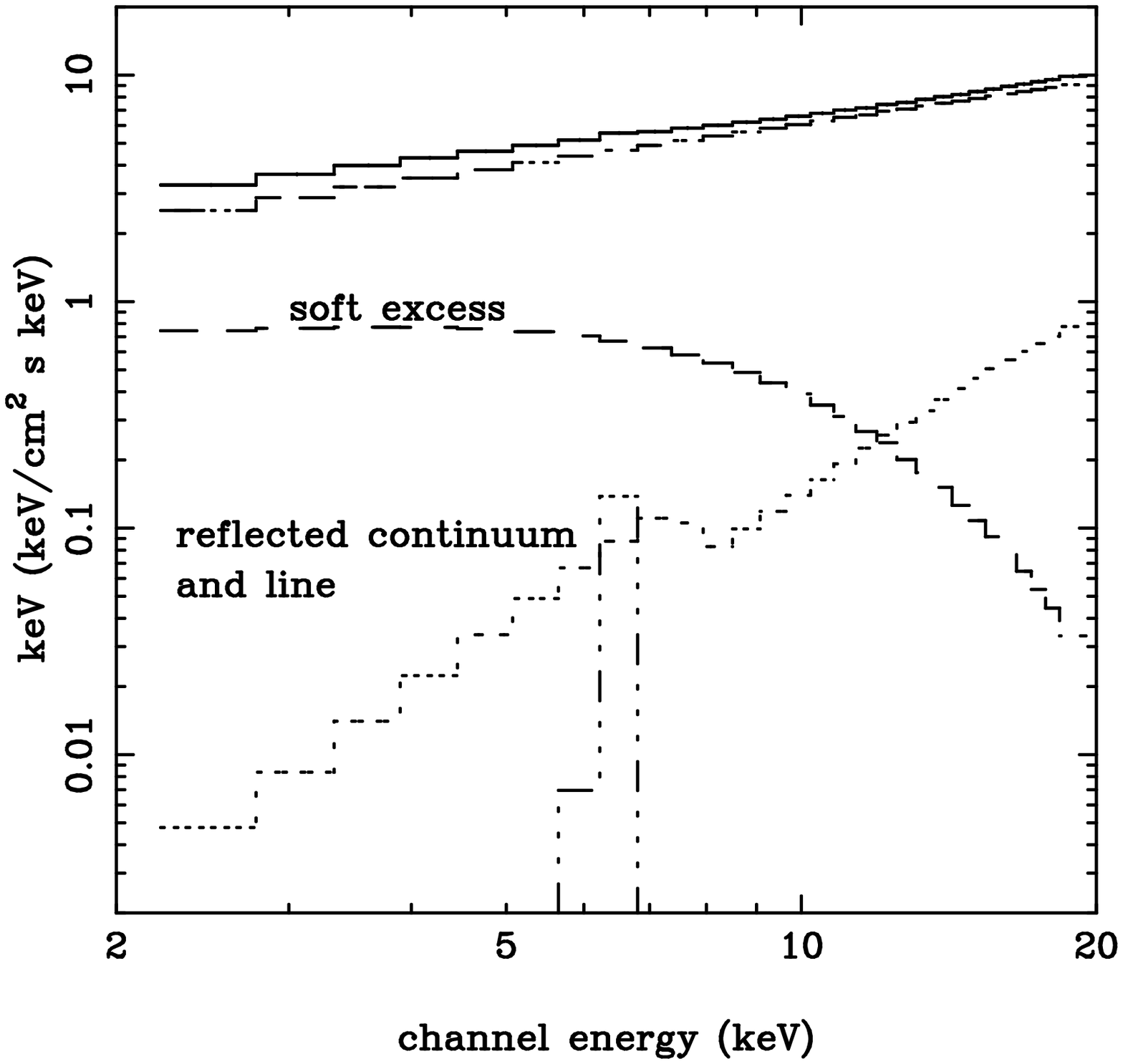,width=.7\textwidth}}
\caption{
The complex ionization model with $\Gamma=1.5$ fit to the single zone
reflector model, together with an additional soft component. The soft
component is strongly required by the ``data'' (compare the fit
statistic in Table 1 and Table 2). It is used to model the excess
reflected flux at low energies which cannot be matched by the fairly
low ionization reflector required by the single zone ionization models
to match the iron edge and rise to higher energies.}
\label{fig:4}
\end{figure*}

\section{Discussion}\label{sect:discussion}

\subsection{Small covering fraction}

The classic ionization instability can cause a dramatic transition in
the vertical structure of an X--ray illuminated accretion disk, with
an extremely ionized skin overlying low ionization
material. With flat spectrum illumination the Compton temperature of
the skin is high enough that even iron is completely stripped, so it
only scatters the incident power law radiation and does not create any
atomic features in the reflected spectrum (such as emission lines and
photo--absorption edges). The only observable reflected signature then
comes from the material below the ionization instability, which has
low--to--moderate ionization. The instability suppresses any signature
of H-- and He--like iron, which was the key argument used by Done \&
\zycki\ (1999) against substantial ionization of the disk. Where the
optical depth in the ionized skin is $\sim 1$ then the observed
reflected spectrum and line derived from single zone ionization models
is suppressed by about a factor 3--4, as observed in the flattest AGN
and GBHC. Thus, the geometry in which the disk covers half the sky as
seen from the X--ray source (e.g., magnetic flares above a disk) {\em
can} be consistent with the observed data because the reflected
spectra computed with hydrostatic balance models {\em appear} to have
a small reflection fraction when fitted with the current single--zone
models. However, more detailed studies of this issue, with radial
stratification of the skin properties, and with effects of possible
X--ray induced evaporation (which is important for magnetic flares, see
NKK) is needed to resolve the issue truly self--consistently.

We estimate that Thomson depth of the skin should be in the range of
unity to few to explain spectra typical of hard state of GBHCs. At
first glance, this seems to be a rather narrow range in parameter
space, making the model appear contrived. However, the dependence of
$\tau_h$ on $\fx$ becomes very weak once $\tau_h\simgt 1$, because the
incident X--rays are efficiently down--scattered in the skin. As
discussed by Kallman \& White (1989) and Nayakshin (2000a), it is then
very difficult to produce $\tau_h\simgt\tau_*\sim $ a few even with
arbitrarily large values of $\fx$.  Therefore, the range of
``acceptable'' values of Thomson depths, $1\simlt \tau_h\simlt $few,
may actually correspond to a change in the X--ray ionizing flux by
several orders of magnitude, and hence the model may be quite robust
in the parameter space.

\subsection{Covering fraction -- spectral index -- ionization correlation}

We find a strong apparent correlation between the inferred intrinsic
spectral index and the amount of reflection and its ionization (see
Fig. 2 and Table 1). The gas pressure at which the transition from the
hot to the cold material occurs, $P_t$, scales with the Compton
temperature as $P_t \propto T_1^{3/2}$ (see eq. 4 in Nayakshin
2000a). Therefore, the cold layer lies closer to the top of the skin
for larger $\Gamma$ for a given illuminating flux.  In our numerical
simulations the Thomson depth of the skin changes from $\tau_h \simeq
1.0$ for $\Gamma = 1.5$ to $\tau_h
\simeq 0.2$ for $\Gamma =2.5$ (see Fig. 1). The thicker the skin, the
more it masks the cold material below it (e.g., see NKK), and so the
correlation between $\Gamma$ and $\Omega/2\pi$ that we found in this
paper is to be expected.

The correlation of the inferred ionization state ($\xi$) with $\Gamma$
comes about as a consequence of changes in $\tau_h$ and the actual
ionization state of the skin with changes in the spectral index. In
particular, when $\Gamma$ increases, the Compton temperature, and
hence the skin's temperature, decrease (see Nayakshin \& Kallman 2000
on this). For $\Gamma\simgt 2$, the skin is highly ionized, rather
than extremely ionized, and so it produces strong features from H-- and
He--like iron. However, its optical depth is not very large for the
parameters chosen in this paper (see Fig. 1). Hence, there is a
substantial contribution to the reflected spectrum from the low to
moderate ionization material under the ionized skin. The ``observed''
ionization state is a compromise between the optically thin, ionized
skin and the low ionization material below, which then leads to the
inferred correlation of $\xi$ and $\Gamma$.

It is also interesting to note that a larger Thomson depth of the
ionized skin leads to larger integrated X--ray albedo, $a$. The thermal
flux resulting from reprocessing of the incident X--ray flux is $F_{\rm
repr} = (1-a)\, \fx$. Thus, the thicker the ionized skin, the less
reprocessed radiation will emerge, which means that Compton cooling of
the magnetic active regions above the disk will be reduced. This is
why Nayakshin (1999b) and Nayakshin \& Dove (2000) find that larger
values of $\tau_h$ lead to hotter coronae above the disk, i.e.,
smaller values of $\Gamma$ if the optical depth of the corona is
fixed. In other words, a complete self--consistent calculation, where
$\Gamma$ is not a fixed parameter, but it is rather found via
balancing energy heating and cooling in the corona would produce {\em
even stronger} apparent correlation between $\Gamma$ and $\Omega/2\pi$
than the one we found here.  These findings appear to be very relevant
for the recently found correlations between $\Gamma$ and $\Omega/2\pi$
for a sample of AGN and GBHCs, as well as for some individual sources
(e.g., Zdziarski, et al. 1999).

Note also that the correlation between $\xi$ and $\Gamma$ inferred in
our models suggest that sources with steeper X--ray spectra should
produce spectra that {\em appear} to be more ionized than those of
sources with harder spectra. This statement is only correct in the
statistical sense, though. Namely, if for every $\Gamma$ there is a
range of illumination parameters $A$, then at a given $A$, there will
be a correlation of $\xi$ and $\Gamma$ similar to that shown in
Fig. 1. Folded with a range of $A$ for every $\Gamma$, one still
expects to see a similar correlation, probably somewhat weaker than
the one we found here.  This theoretical expectation does indeed
appear to match the behavior of the (steep spectrum) Narrow Line
Seyfert 1's, which have a large amount of ionized reflection (NLS1 --
Pounds, Done \& Osborne 1995; Vaughan et al. 1999). This supports the
picture where the difference between NLS1s and normal Seyfert 1s is
that NLS1s accrete at a high fraction of the Eddington limit (Pounds
et al. 1995), as the steep spectrum gives a low Compton temperature
and so reflection from highly ionized as opposed to extremely ionized
material. However, this correlation does {\em not} hold for
MCG--6--30--15, which has a steep spectrum together with a large
amount of apparently {\em neutral} reflection (Lee et al 1999;
Zdziarski et al 1999), although the variability of the line may
indicate that ionization effects are important (Reynolds 2000).

\subsection{Multi--radius reflection spectra}

Clearly the models used here are incomplete. The optical depth of the
ionized layer changes as a function of radius, and relativistic
smearing effects should also be included. However, it is clearly
necessary to develop  a basic understanding of how the complex
ionization models differ from simplified single zone ones for a
specific radius with no relativistic effects. Having understood that,
one can move to the reflected spectra for complete disks, which we
plan to do in future work. 

Based on physical intuition and some preliminary work, we can
speculate on the full disk spectra. NKK and Nayakshin \& Kallman
(2000) show that if the illuminating X--ray flux, $F_x$, exceeds that
of the locally produced disk thermal emission, $F_d$, and the X--ray
spectrum is relatively hard, i.e., $\Gamma\simlt 2$, then the skin is
nearly completely ionized. The depth of the completely ionized skin
decreases with increasing radius in the disk (see Nayakshin
2000a). Hence, larger radii contribute relatively more of the atomic
features (e.g., K$\alpha$ line and iron edges) to the overall
spectrum, while smaller radii contribute less. If $\tau_h >>1$ in the
inner disk then the extremely relativistic reflected signature will be
absent from the data. A low ionization reflection signature is only
seen from radii larger than the point at which $\tau_h \sim 1$. This
would suppress the broadest components of the iron line, so could make
the derived line profile look {\em qualitatively} like that from a
truncated disk (see Fig. 2b in Nayakshin 2000b). One still hopes that
detailed and direct spectral fitting, combined with future more
sensitive observations, will allow us to discriminate between the
physically truncated and completely ionized disks.

\subsection{Soft X--ray excess}

The key element in the complex ionization models is that for hard
X--ray spectra the ionization instability results in a sharp
discontinuity between the extremely ionized (invisible) skin and the
low ionization material which gives a clearly identifiable reflection
signature. Given further complications that can arise due to
relativistic broadening in the disk, it is quite likely that the {\em
full} X--ray illuminated disks have a reflected continuum which can
mimic that of a truncated disk. Can we then distinguish between these
two models?  We have shown that there are {\em observational} tests.
Firstly broad bandpass, high signal--to--noise data should show the
deficiencies in the {\tt pexriv} model fit (see Figure 3).  Such data
currently exist only for the Galactic Black Hole Candidates, with Cyg
X--1 having the best determined spectrum (Gierlinski et al. 1997; di
Salvo et al. 2000).  These clearly show that the reflected spectrum
from single ionization state material illuminated by a Comptonized
continuum is insufficient to describe the observed spectral curvature.

Further, even in the 2--20 keV bandpass there is an observable
difference between the single zone and complex ionization reflection
models. The complex ionization disks predict the existence of a
highly ionized reflected skin which contributes to the spectrum at low
energies. Physically, since there is no photo--absorption opacity left
in the skin, photons of all energies are scattered (reflected) rather
than being absorbed in the skin. 

The low ionization reflection component has
negligible flux at low photon energies (see, e.g., Basko et al. 1974,
Lightman \& White 1988), so that spectrum from complex ionization
models yields a soft excess as compared with the single--zone models,
which can be modeled by a broad spectral component at low energies.
An additional and significant source of the soft X--ray excess is the
bremsstrahlung emission in the skin, and Compton down--scattered
reflected flux. According to the analytical
theory of the Compton--heated completely ionized skin (Krolik et
al. 1981; Nayakshin 2000), cooling due to bremsstrahlung emission
constitutes between $\sim$ zero (at the top of the skin) and $2/3$ of
the total cooling rate (on the bottom). The characteristic temperature
at which the Compton--heated branch ceases to exist is of order a few
hundred eV to a few keV (see Fig. 1), so that the soft X--ray excess
will be rolling over at similar energies, consistent with our results
here.

There is a considerable evidence for a similar soft excess in the
Galactic Black Hole Candidates. Cyg X--1 plainly shows such a
component in ASCA (Ebisawa et al., 1996) and SAX (di Salvo et al.,
2000) data, and this is probably the reason why the RXTE results
include a blackbody at $\sim 1$ keV rather than a reflected component
(Dove et al 1997). Figure 5a shows the PCA RXTE data of Cyg X--1
(dataset 10241, as used by Dove et al., 1997) fit to a simple single
zone ionization model ($\chi^2_\nu=19.1/38$), while Figure 5b shows
the resulting fit including soft component
($\chi^2_\nu=11.5/35$). The spectrum is much better
described with the inclusion of the soft component. The resulting
spectral curvature is detected at $\ge 99$ per cent confidence on an
$F=(\Delta\chi^2/\delta\nu)/\chi^2_\nu = 7.7$ test, and these real data
look remarkably like the synthesized model spectra shown in Figure 4.

\begin{figure*}
\begin{minipage}[b]{.46\textwidth}
\centerline{\psfig{file=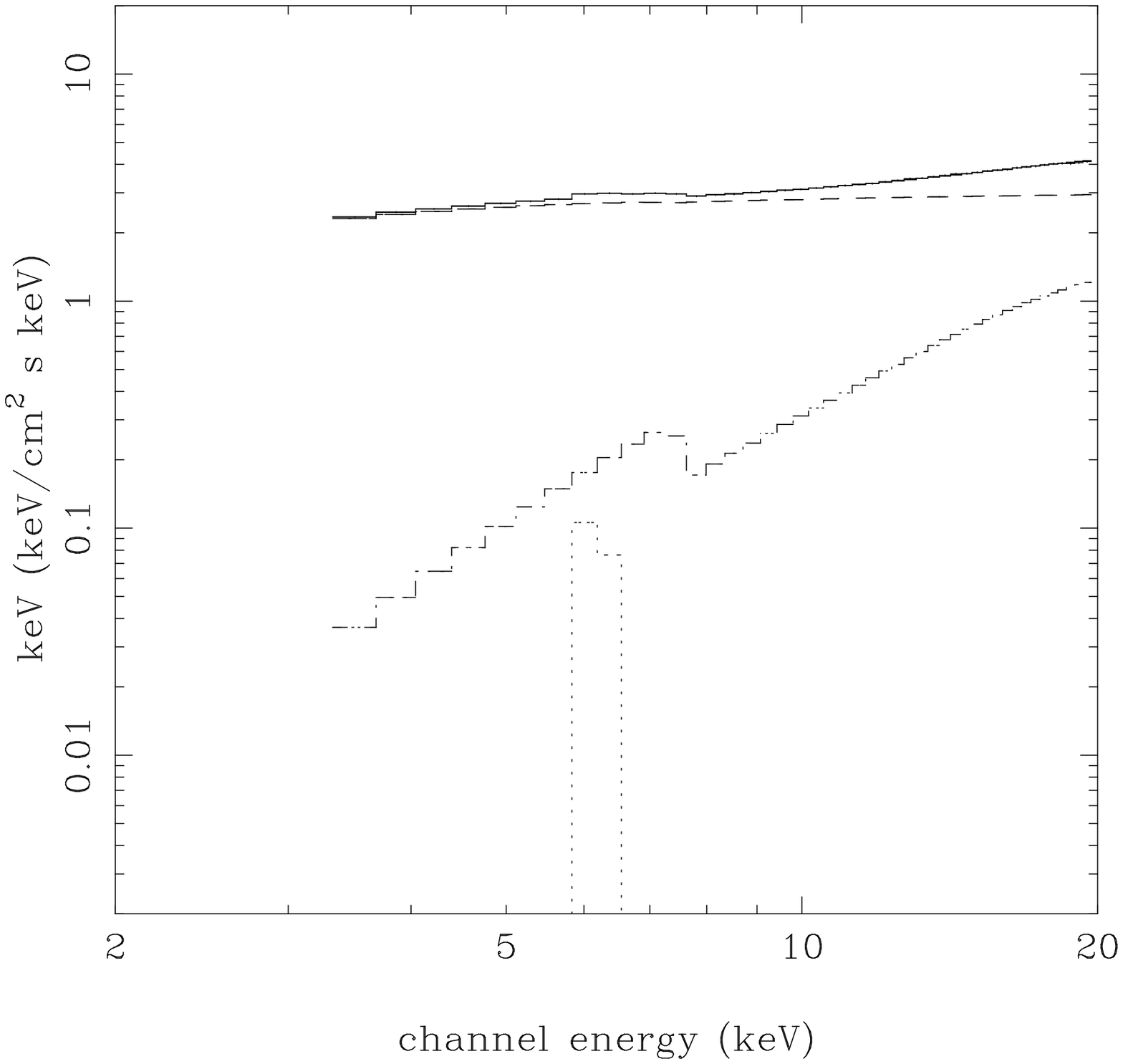,width=\textwidth}}
\end{minipage}\hfill
\begin{minipage}[b]{.46\textwidth}
{\psfig{file=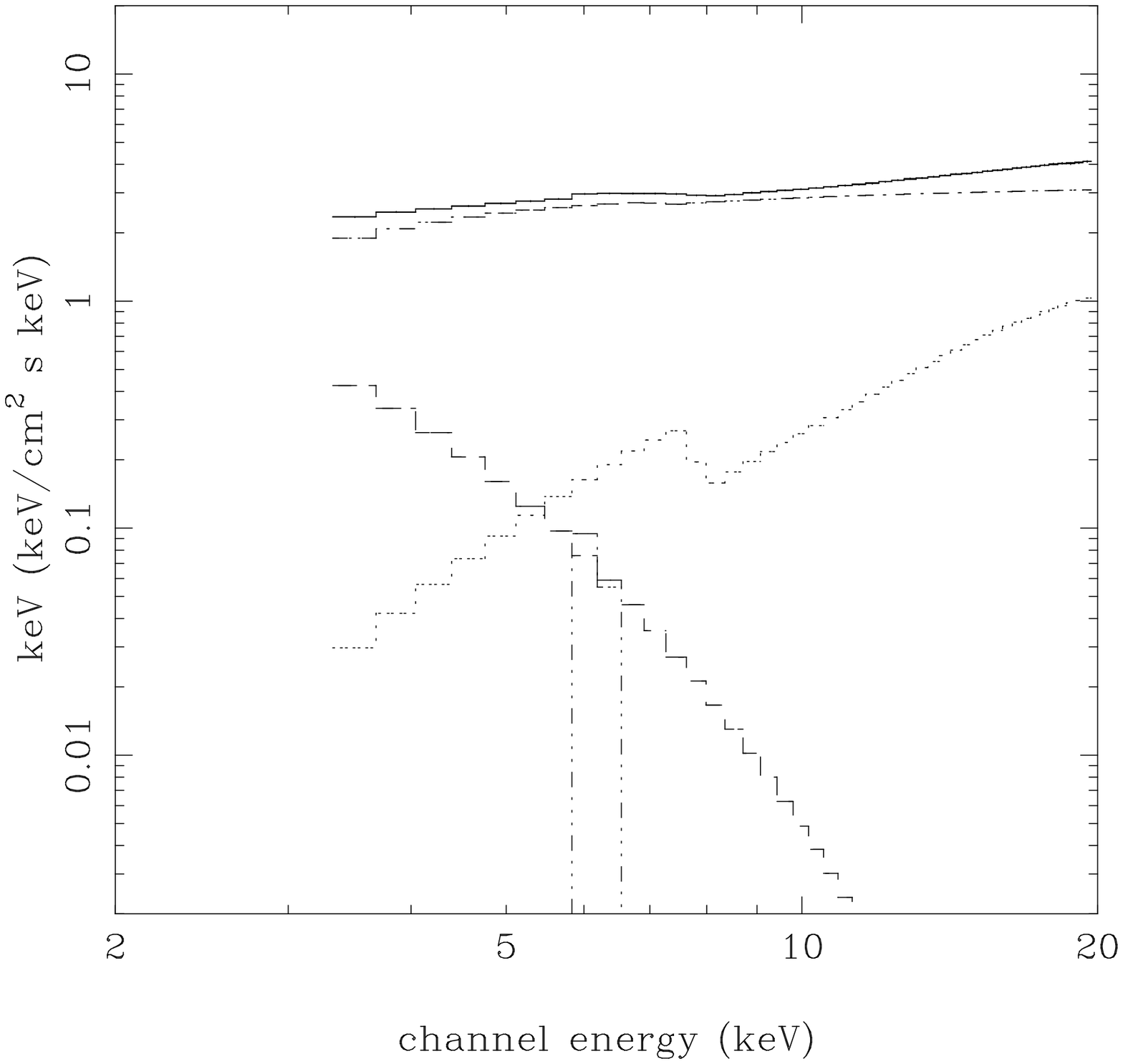,width=\textwidth}}
\end{minipage}
\caption{
The single zone ionization model fit to the {\em observed} RXTE PCA
spectrum of Cyg X--1. This instrument has similar energy resolution to
GINGA, and 0.5 per cent systematic errors are included on the top
layer data from PCA detectors 0 and 1. Panel (b) shows the fit
including a soft component.}
\label{fig:5}
\end{figure*}

Clearly this can be interpreted as strong evidence for the existence
of complex ionization structure in the accretion disk. However, this
is not unambiguous as the GBHC have other ways to produce spectral
curvature since the seed photons from the disk are so close to the
energy range observed.  A truncated disk with hot inner flow produces
a Comptonized spectrum where the first few scattering orders are
anisotropic, since the illuminating disk photons are anisotropic. This
produces a soft excess of about 10 per cent of the isotropic spectrum,
i.e.  similar in size to that expected from the complex ionization
reflected spectrum (P.T. \zycki, private communication).
Additionally, the transition between the cool truncated disk and hot
inner flow is unlikely to be completely sharp, and there may be a
further soft component from heated disk material.  These would both
affect the GBHC spectra more than those from AGN, so perhaps the true
test of whether the disk has complex ionization structure or is
truncated is whether the AGN with flat spectra are significantly
better fit by including a soft excess component.  However, even with
the AGN there are ambiguities. One of the most promising models for
the origin of the variability behavior of AGN and GBHC has spectral
evolution of the Comptonized component from individual flares. This
results in a time averaged spectrum which is significantly curved,
similar to the curvature seen here in Cyg X--1 (Poutanen \& Fabian
1999). We caution that deconvolving the reflected spectrum from more
realistic, complex continuum models will have to be done before we can
observationally test whether the truncated disk or complex ionization
models give a better description of the data.

\section{Summary}

In this paper, we have shown that reflected X--ray spectra from models
which include hydrostatic balance for the illuminated gas have very
different properties than single zone ionization models. In the former
models, due to the thermal ionization instability, the vertical disk
structure is almost discontinuous, with a dramatic transition from an
extremely ionized skin on the top to a low--to--moderate ionization
material below it. The reflection signature is then a composite of 
a power law like spectrum from the extremely ionized material, and a
low--to--moderate ionization reflection siganture from the material 
below the skin. The constant density reflection models
do not have such a discontinuous behavior. As a result,
when the spectra of hydrostatic balance models are fit with simpler,
single ionization zone models (Done et al. 1992; Magdziarz \&
Zdziarski 1995), the physical parameters of the reflector can be
miscalculated by a large factor. 
The single zone models fit to the low--to--moderate reflection
component, which is weak even with respect to the incident power law
due to scattering in the ionized skin. But the ionized skin reflection 
cannot be reproduced in the single zone models, so this power law like
reflected component is misidentified with the incident power law.
Since the reflected fraction $\Omega$ is proportional to
the ratio of the inferred reflected spectrum to that of the
illuminating, then this further dilutes the ``observed'' reflected
fraction.

This implies that the values of $\Omega/2\pi<1$ often seen in
GBHC and AGN may be an artifact of the constant density reflection
models used, and need not necessarily mean that the cold SS disk is
truncated at some distance $R > 3 R_s$. Further, our results also
imply that a low value of ionization parameter (and ``neutral'' line at
6.4 keV), as found by the single zone ionization models, {\em does
not} prove that the material is actually cold. On the contrary,
the upper layer of the disk may be completely ionized and obscure the
cold material below. These findings are in agreement with earlier
suggestions of Nayakshin (1999b) and Ross, Fabian \& Young (1999).

In addition, we find an apparent correlation between the power law
index $\Gamma$ of the X--ray continuum and the reflection
fraction. This correlation is similar to the one reported by Zdziarski
et al., (1999) for AGN and GBHC. We produce the correlation using only the
physical nature of the ionization instability for X--ray illuminated
gas in hydrostatic balance. This should operate in {\em all} models
where the accretion disk is illuminated by an X--ray source.  In
particular, other ways to produce the correlation of the amount of
reflection with $\Gamma$ such as a moving inner disk edge
(e.g. Poutanen 1998) or with anisotropic illumination by magnetic
flares (Beloborodov 1999a,b) will produce even stronger correlations
with the inclusion of the ionization physics. The {\em only} situation
where the instability is unimportant for the observed reflected
spectrum is if the optical depth of the hot layer is always very much
less than unity. However, one clear prediction of the ionization instability
is that the steep spectrum AGN should have strongly ionized reflected
spectra, which does indeed seem to be the case for the narrow line
Seyfert 1 class of AGN (although not for the archetypal extreme
relativistic disk object MCG--6--30--15). Thus it seems likely that
the ionization instability {\em is} important in determining the
observed reflected spectrum, although it may operate in conjunction
with anisotropic illumination or a moving disk edge.

We show that reflection from the ionized skin produces a soft X--ray
continuum which can be modeled as a broad soft X--ray excess, similar
to that seen in data from Cyg X--1. Taken together, our findings revive
models for magnetic flares above an untruncated accretion disk in AGN
and GBHCs which seemed to be ruled out by observations of the small
solid angle of the reflector (see, e.g., the discussion in Done \&
\zycki\ 1999). It seems likely that radial gradients in the ionization
structure should be able to suppress the extreme relativistic iron
line profile from the inner disk, and so lead to the observed apparent
truncation of the disk in the low/hard state GBHC and in some
AGN. However, this should be confirmed by further detailed modeling
involving complete disk spectra and relativistic smearing.

Finally, another difficulty in unambiguously determining between
complex ionization structure on a disk which extends down to the last
stable orbit, and a truncated disk, is that the underlying
illuminating continuum shape is not well understood. Continuum
curvature can mimic the ``soft excess'' signature of complex ionization
structure, and is predicted in the truncated disk geometry, and by
models of the variability. Sadly, the observational constraints on the
geometry of the accretion flow around the black hole in Cyg~X--1 still
cannot distinguish between disk/coronae and ADAF models for the X--ray
emission. On the other hand, new data and complete disk calculations
may present better constraints, allowing these very different
geometries (and accretion theories) to be distinguished.

\section{ACKNOWLEDGEMENTS}

SN acknowledges support from NRC Research Associateship. The authors
thank Demos Kazanas, Piotr \zycki\ and Andrei Beloborodov for useful
discussions.

\end{document}